\documentclass[]{rsos}

\usepackage[show]{chato-notes} 

\usepackage{array}
\newcolumntype{L}[1]{>{\raggedright\let\newline\\\arraybackslash\hspace{0pt}}m{#1}}
\newcolumntype{C}[1]{>{\centering\let\newline\\\arraybackslash\hspace{0pt}}m{#1}}
\newcolumntype{R}[1]{>{\raggedleft\let\newline\\\arraybackslash\hspace{0pt}}m{#1}}

\usepackage{blindtext}
\usepackage{helvet}
\usepackage{courier}
\usepackage{graphicx}
\usepackage{subfig}

\usepackage{balance}
\usepackage{soul,color}
\soulregister\cite7
\soulregister\ref7
\soulregister\pageref7
\begin{document}

\title{Dynamics and Biases of Online Attention: The Case of Aircraft Crashes}

\author{
Ruth  Garc\'{i}a-Gavilanes, Milena Tsvetkova and Taha Yasseri}

\address{Oxford Internet Institute, University of Oxford, U.K}

\subject{behaviour, complexity, human-computer interaction}


\corres{Taha Yasseri\\
\email{taha.yasseri@oii.ox.ac.uk}}

\begin{abstract}
The Internet not only has changed the dynamics of our collective attention, but also through the transactional log of online activities, provides us with the opportunity to study attention dynamics at scale. In this paper, we particularly study attention to aircraft incidents and accidents using Wikipedia transactional data in two different language editions, English and Spanish. We study both the editorial activities on and the viewership of the articles about airline crashes. We analyse how the level of attention is influenced by different parameters such as number of deaths, airline region, and event locale and date. We find evidence that the attention given by Wikipedia editors to pre-Wikipedia aircraft incidents and accidents depends on the region of the airline for both English and Spanish editions. North American airline companies receive  more prompt coverage in  English Wikipedia. We also observe that the attention given by Wikipedia visitors is influenced by the airline region but only for events with high number of deaths. Finally we show that the rate and time span of the decay of attention is independent of the number of deaths and a fast decay within about a week seems to be universal. We discuss the implications of these findings in the context of attention bias.

\end{abstract}

\maketitle

%
%
%
%


\section{Introduction}

The Internet has drastically changed the flow of information in our society. Online technologies enable us to have direct access to much of the world's established knowledge through services such as Wikipedia and to informal 
user-generated content through social media. There is no theoretical limit to the information bandwidth on the Internet but human attention has its own limits. Public attention to emerging topics decays over
time or suffers the so called memory buoyancy from users, which is a metaphor of information objects sinking down in the digital memory with decreasing importance and usage, increasing their distance to the user \cite{Kanhabua2013}. 

Nowadays, the online footprints of users have rendered the level of attention given to new and past events and its decay an observable phenomenon. The digital nature of Internet-based technologies enables us to analyse the variances of attention at a scale and with an accuracy that have not been feasible in relation to other communication technologies. Researchers have used logs generated by online users' activities such as tweets, search queries, and web navigation paths to cover a wide range of topics on attention. For example, Lehmann~\emph{et al.}~\cite{Lehmann2012} characterize attention by analysing the time-series of tweets with popular tags from a data set of 130 million tweets from 6.1 million users and found four clusters based on dynamics, semantics, and information spread. Yeung~\emph{et al.}~\cite{Yeung2011} focus on how events are remembered for specific years by looking at temporal expressions in the text of 2.4 million articles in English from Google news archive; they find more references to more recent events. Other studies have concentrated on attention decay. Wu and Huberman \cite{wu2007} discover a very short time span of collective attention with regard to news items on the digg.com linksharing website. Simkin and Roychowdhury~\cite{Simkin2012} study blogs and news from more than 100 websites and find that decay in accessibility is due to aspects of visibility such as link positioning and attractiveness. Researchers have also linked online attention to more practical matters, from predicting election outcomes \cite{YasseriB14} and detecting memory patterns in human activities~\cite{SingerDet2014}, all the way to analysing trading behaviour in financial markets \cite{PreisQuant2013} or the appropriate time when to publish news to gain more attention~\cite{Subasic2013}.

While several aspects of online attention increase and decay have been fairly well investigated, much less is known about how geography, event impact, and differences across populations with different languages affect attention.
Thus, the question whether online technologies have improved or worsen the fairness  and equality with which news are released to the public, influencing their attention, is still open. 
The question is particularly important  to investigate with regard to high impact events such as the terrorist attacks in Paris and Beirut in November 2015. It was reported \cite{Roy2015} that only 11\% of the top media outlets covered the Beirut attacks in the first 24 hours in comparison to 51\% for Paris. Furthermore, user attention  for the Beirut bombings  within the first hour was only 5\% of what Paris achieved within the same time period in spite of the Paris attacks starting almost 15 hours after Beirut. 
What determines what is covered by the media and when?  What determines the level of public attention to new events? Does the decay of public attention varies depending on the event?  In this paper, we answer these questions at scale by analysing editorial and traffic information on a set of articles in two different language editions of Wikipedia. We study how events are covered, what aspects determine attention to them, how attention decays, and whether there are differences between languages. Focusing on depth rather than breadth, we limit our analyses to one specific type of event---aircraft incidents and accidents---and to the two most popular Wikipedia language editions by number of active users---English and Spanish. 

Wikipedia is a unique resource to study collective attention. Written and edited by volunteers from all around the world, it has become the number one source of online information in many languages, with close to 40 Million articles in around 300 language editions (and counting) and with open access to logs and metadata. There is a high correlation between search volume on Google and visits to the Wikipedia articles related to the search keywords \cite{Ratkiewicz2010,Yoshida2015}. This indicates that Wikipedia traffic data is a reliable reflection of web users' behaviour in general. The high response rate and pace of coverage in Wikipedia in relation to breaking news \cite{Althoff2013,KeeganHow2011} is another feature that makes Wikipedia a good research platform to address questions related to collective attention. For instance, researchers have analysed Wikipedia edit records to identify and model the most controversial topics in different languages \cite{YasseriThemost2013,iniguez2014modeling}, to study the European food culture \cite{LauferMin2014}, and to highlight entanglement of cultures by ranking historical figures \cite{Young-HoHigh2013}. Wikipedia traffic data has also been used to predict movie box office revenues~\cite{MartonEarly2013}, stock market moves \cite{moat2013}, electoral popularity~\cite{YasseriB15}, and influenza outbreaks~\cite{McIver2014,hickmann2015forecasting}.

To answer our research questions, we develop an automatic system to extract editorial and traffic information on the Wikipedia articles about aircraft incidents and accidents and factual information about the events. By comparing the English and the Spanish Wikipedia, we contribute to this research field in the following ways: 

\begin{itemize}

\item We study the coverage of the events in Wikipedia and its dynamics over time considering the airline region, the event locale, and the number of deaths. 

\item We analyse the role of the airline region and number of deaths on the viewership data to Wikipedia articles. 

\item We model attention decay over time.

\end{itemize}

 We present the results from our study in the next section, after which we continue with discussion and conclude with implications. Details for our data collection and analysis strategy can be found in the last section, Section \ref{sec:matmeth}.

\section{Results}
\label{sec:res}

\begin{figure*}[t]
  \centering
         \includegraphics[width=1\textwidth]{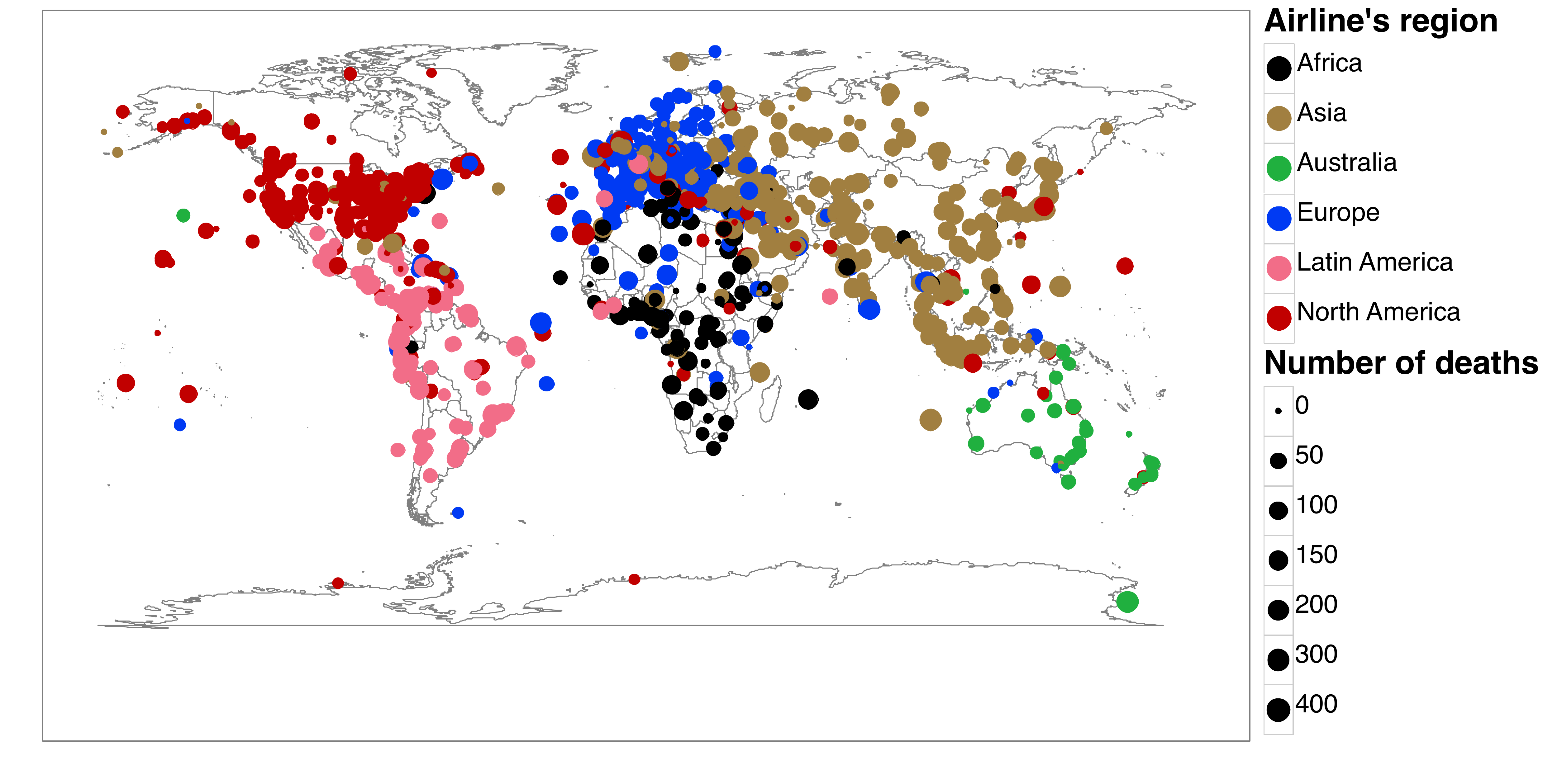}

  \caption{1496 geolocated incidents and accidents since 1897 reported in English Wikipedia. Each dot represents an event. The size of the dots is proportional to the number of reported deaths and the colour codes the location of the operating company.   }\label{fig_world_map}
\end{figure*}

Figure~\ref{fig_world_map} shows a map of all the aircraft incidents and accidents from English Wikipedia coloured according to the airline region, which is where the airline company for the flight is located, and sized according to the number of deaths caused by the event.
For simplicity, we divide the Americas into two regions: North America and Latin America. Latin America includes all countries or territories in the Americas
where Romance languages are spoken as first language (in this case, Spanish, Portuguese, and French) and all Caribbean islands, while North America includes the rest
(i.e., mostly United States and Canada). Furthermore, all headquarters in the EuroAsia region are labeled as Asia (e.g., Russia and Turkey).
We observe that the locales of the events overlap most of the time with the airline regions.

Our results are divided in three sections: the first part deals with the editorial coverage of the events, the second with the immediate collective attention quantified by viewership statistics, and the third with the modelling of attention decay.

\subsection{Editorial Coverage}

Table \ref{table_groundtruth} compares
the number of aircraft accidents and incidents  covered in English and Spanish Wikipedias with cases reported by the Aviation Safety Network (ASN)\footnote{\url{http://aviation-safety.net/statistics/geographical/continents.php}} in different continents.
While ASN provides data  from 1945, excluding military accidents, corporate jets, and hijackings, our dataset includes these
cases and dates back to the year 1897. 
There are 1,081 articles in English Wikipedia that do not have a Spanish equivalent and most of them are about events that 
happened in North America (265),  Asia (261), and  Europe (252). On the other hand, there are 71 articles in Spanish Wikipedia
with no English equivalent and most of them are about events that happened in Latin America (39). 

\begin{table*}[t]
\scriptsize
\centering
\begin{tabular}{llrllrllrl}
  
  \multicolumn{1}{c}{} & \multicolumn{6}{c}{\textbf{ Wikipedia}}&\multicolumn{3}{c}{\textbf{ ASN}}\\ \cline{2-7}
\multicolumn{1}{c}{} & \multicolumn{3}{c}{\textbf{ English}} & \multicolumn{3}{c}{\textbf{ Spanish}} & \multicolumn{3}{c}{}  \\
& Events & \multicolumn{2}{c}{Deaths} & Events & \multicolumn{2}{c}{Deaths} & Events & \multicolumn{2}{c}{Deaths}  \\
Continent &  & avg &  total &  & avg &  total &  & avg & total \\ 
  \hline
Africa & 0.08 & 49 &  5,967 & 0.07 & 58 &  1,981 & 0.10 & 20 & 8,108 \\ 
  Asia & 0.24 & 50 &  17,987 & 0.22 & 61 & 6,618 & 0.17 & 27 & 19,351 \\ 
  Australia & 0.03 & 21  & 873 & 0.01 & 52  & 260 & 0.03 & 12 & 1,448 \\ 
  Europe & 0.22 & 36  & 11,818 & 0.17 & 59  & 4,963 & 0.24 & 23 & 23,423 \\ 
  L. America & 0.08  & 47 & 5,789 & 0.24 & 40  & 4,695 & 0.19 & 16 & 12,942 \\ 
  N. America & 0.23 & 27  & 9,052 & 0.16 & 45  & 3,517 & 0.23 & 13 & 12,958 \\ 
  Others & 0.12 & 45 &  8,353 & 0.13 & 80 &  4,941 & 0.02 & 32 & 2,712 \\ 
  Total & 1,496 & 40 &  59,839 & 488 & 55  & 26,975 & 4,223 & 19 & 80,942 \\ 
   \hline
\end{tabular}
\caption{Breakdown by region of the number of aircraft incidents and accidents covered in Wikipedia compared to the data available at The Aviation Safety Network (ASN) website.
The column \emph{Events} is the ratio with regard to the row \emph{Total}.}\label{table_groundtruth}
\end{table*}

With regard to the number of deaths, the lowest average numbers correspond to  Australia, North America, and Europe respectively for
English Wikipedia, whereas Latin America and North America have the lowest average number of deaths for Spanish Wikipedia. 
This is because some low impact events (many with 0 deaths) that occurred in Australia, North America, and Europe are only included in English Wikipedia
and some low impact events in Latin America are only considered notable in Spanish Wikipedia. With regard to the articles in English that do not have a
Spanish equivalent, the average number of deaths is 39 and  for those that do not have an English equivalent the average is 12. These numbers indicate that
the articles in Spanish without an English equivalent are low impact events concentrated in Latin America.

\begin{figure}[t]
                 \centering
                        \includegraphics[width=1\textwidth]{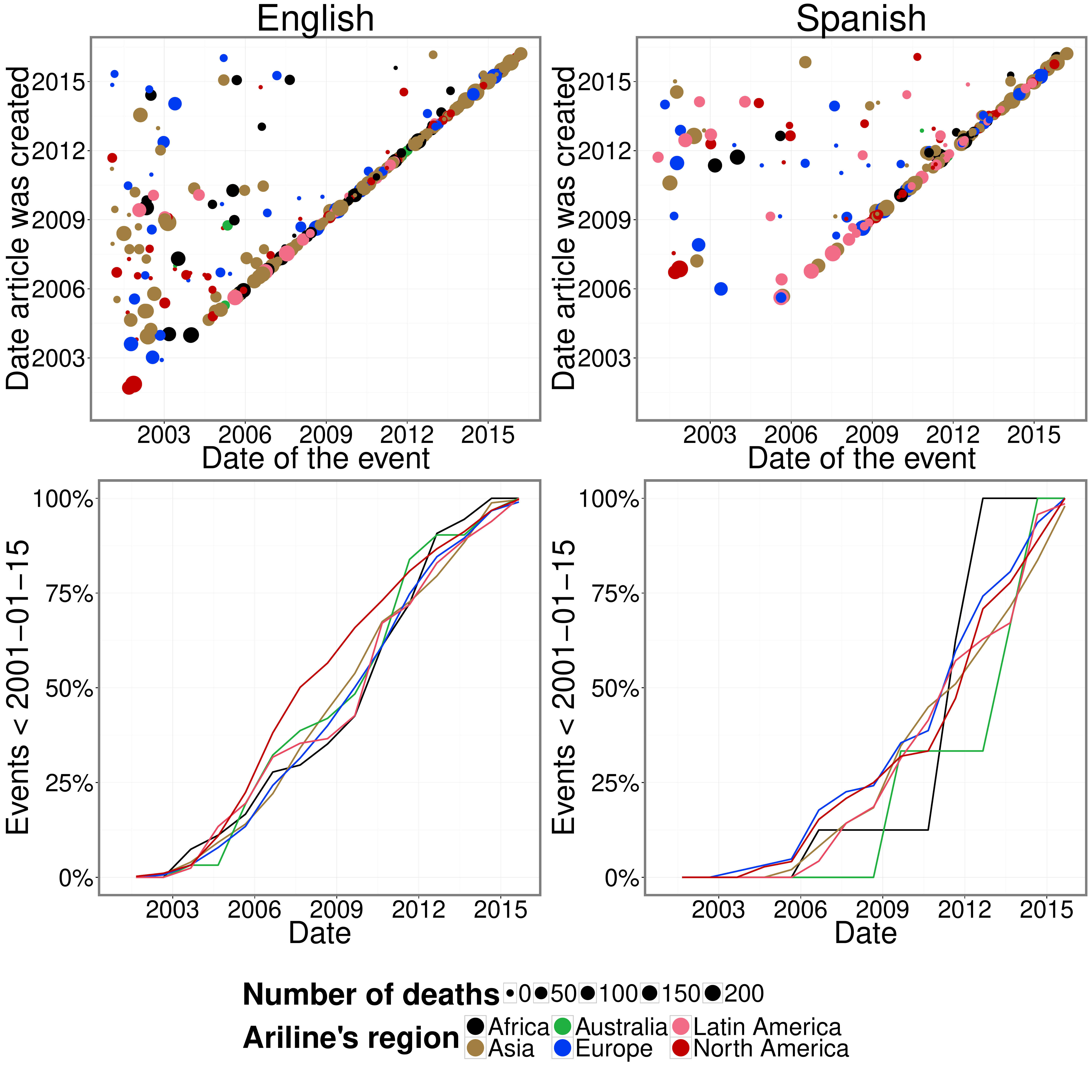}

   \caption{Coverage of aircraft incidents and accidents in the English and Spanish Wikipedia: the upper panels show the lag between the occurrence of the event and the creation of the corresponding article in Wikipedia for post-Wikipedia events, the lower panels show the corresponding percentage of covered pre-Wikipedia events in time.}\label{fig_date_comparison}
 \end{figure}

We also investigate the time lag between the occurrence of the event and the creation of the corresponding Wikipedia article.
Our dataset contains articles about events that happened before and after Wikipedia was launched (see Fig.\ref{fig_events_per_year} in the Appendix).
Post-Wikipedia events (399 for English and 224 for Spanish) are shown on the upper row panels of Figure~\ref{fig_date_comparison}, where the horizontal and vertical axes show the time of the occurrence of the event and the creation of the corresponding Wikipedia page respectively. The convergence of the data points towards 
the diagonal line indicates that the community of Wikipedia editors reacts increasingly fast to this kind of events. 
English Wikipedia has been faster at covering events since the diagonal trend starts earlier. A possible explanation is the larger number of users in English Wikipedia compared with the Spanish version.

The lower row panels of Figure~\ref{fig_date_comparison} show the coverage of the pre-Wikipedia events. The colour of the curve corresponds to
 the airline's region and the x-axis shows the year of the Wikipedia page creation. For English Wikipedia (1,078 cases) a quicker coverage of North American events is
 evident. African, Australian, and South American events exhibit sharp increases as the addition of these articles was concentrated in specific periods. On the other hand, Spanish
 Wikipedia (264 cases) shows a slightly faster coverage for events related to European companies  with sharp jumps for African and Australian companies (there are only 34 and 
 5 cases respectively). Most importantly, however, not only did English Wikipedia cover more pre-Wikipedia events, but it also did it faster. Again, this can be explained considering the arger size of the editorial community of English Wikipedia.

\subsection{Immediate attention}\label{sec:maxattention}

Now we turn to the viewership data. 
To capture the immediate attention to an event right after its occurrence, we choose the articles that were created up to 3 days after the event
and extract the maximum number of views within 7 days after the page was created (see figure~\ref{fig_example_fitting} for an example). We discuss the choice of 7 days in section \ref{sec:attdyn}.

\begin{figure}[t]
\centering
       \includegraphics[width=1\textwidth]{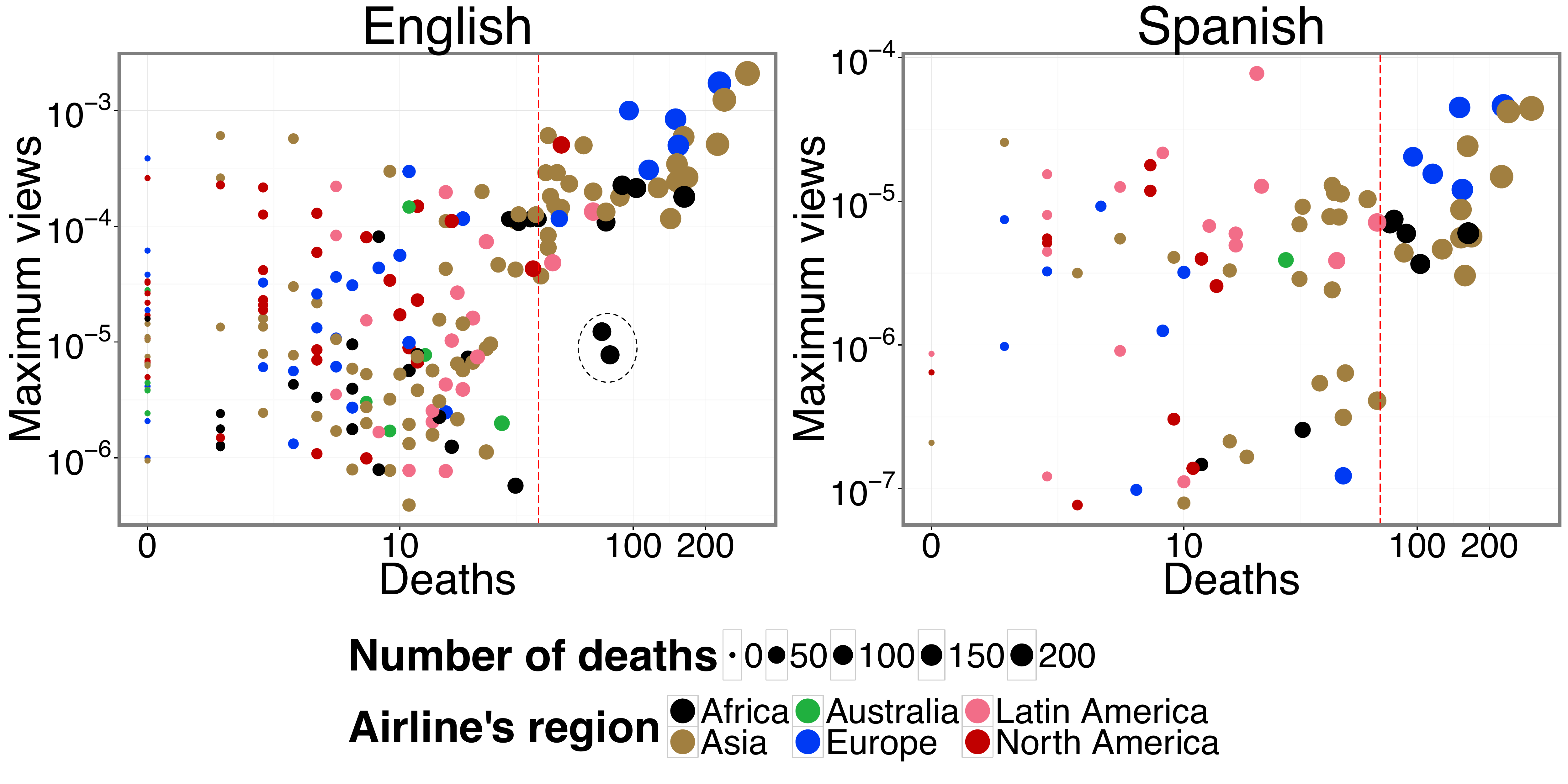}

  \caption{Normalized maximum number of page views versus the number of deaths of each event, both on log-scale for English (En) and Spanish (Sp) Wikipedia. The two outliers in the left panel are removed from the analysis.} \label{fig_regressions}

\end{figure}

A baseline hypothesis would be that the larger the number of 
deaths the event caused, the more attention it attracts. However, this is not always the case; attention is driven by other factors such as media coverage, location, people involved, etc.
This is reflected in Figure \ref{fig_regressions}. The plot shows the normalized maximum daily views versus the number of deaths in log scale for the English and Spanish Wikipedias.

In English Wikipedia, we have identified two regimes: low-impact events ($< 40$ deaths), where there is no correlation between impact and attention, and high-impact events ($\geqslant 40$ deaths), where the maximum number of daily page views increases proportionally to the event impact with $r=0.71$, $p<0.001$. To separate these two regimes, we used visual inspection to accommodate the largest empty square on the lower-right region of the diagram. Regardless of the high correlation of this region,  impact does not always reflect attention: the plot  shows two African outliers with less attention than expected from the overall trend. In Spanish Wikipedia, the separation of the two phases at around 70 deaths is less evident but still exists. The correlation in the high impact regime is $r=0.67$, $p<0.005$. Also note that in the high impact regime, the level of attention increases almost quadratically with the number of deaths. However, we hesitate fitting a function here due to the small number of data points. 

To analyse the importance of the airline's region  and number of deaths on level of attention, we use linear
regression models. We have removed the two outlier events from the English sample shown in Figure \ref{fig_regressions}.
 We then model all the data points using a simple linear model considering the number of deaths as the only parameter (see Table~\ref{table_regression}). In the English case, 
the number of deaths alone can only explain around 22\% of the variation in the level of the immediate attention. 
If we add the airline region as a categorical variable using Africa as the reference category, 
we increase the explanatory power to 28\%. Here, we observe that events related to North American companies attract more views than companies from other regions ($\beta_{1}=1.67$).
On the other hand, Latin American companies play the same role in Spanish Wikipedia ($\beta_{1}=1.68$).

\begin{table*}[t]
\centering
\begin{tabular}{lrrrrrrrr}
\multicolumn{1}{c}{}&\multicolumn{8}{c}{ \textbf{ All events}}\\\cline{2-9}
\multicolumn{1}{c}{}&\multicolumn{4}{c}{ \textbf{ English (n = 204)}}&\multicolumn{4}{c}{ \textbf{Spanish (n = 80)}}   \\
&\multicolumn{1}{c}{$\beta_{1}$} && \multicolumn{1}{c}{$\beta_{2}$}& &\multicolumn{1}{c}{$\beta_{1}$} && \multicolumn{1}{c}{$\beta_{2}$}& \\\hline
  \hline
Intercept & -12.18 & *** & -13.19 & *** & -13.89 & *** & -15.24 & *** \\ 
  Deaths & 0.61 & *** & 0.69 & *** & 0.41 & **  & 0.53 & *** \\ 
  Asia &  &  & 0.79 & *  &  &  & 0.7 &   \\ 
  Australia &  &  & 0.22 &   &  &  & 0.99 &   \\ 
  Europe &  &  & 1.42 & **  &  &  & 1.21 &   \\ 
  Latin America &  &  & 0.23 &   &  &  & 1.68 & *  \\ 
  North America &  &  & 1.67 & *** &  &  & 0.96 &   \\ 
  Adj. $R^2$ & 0.22 & *** & 0.28 & *** & 0.11 & **  & 0.12 & *  \\ 

\hline
  \multicolumn{1}{c}{}&\multicolumn{8}{c}{ \textbf{ Low-impact}}\\\cline{2-9}
\multicolumn{1}{c}{}&\multicolumn{4}{c}{ \textbf{ English (n = 166)}}&\multicolumn{4}{c}{ \textbf{Spanish (n = 60)}}   \\
&\multicolumn{1}{c}{$\beta_{1}$} && \multicolumn{1}{c}{$\beta_{2}$}& &\multicolumn{1}{c}{$\beta_{1}$} && \multicolumn{1}{c}{$\beta_{2}$}& \\\hline
Intercept & -11.44 & *** & -12.27 & *** & -13.3 & *** & -15.87 & *** \\ 
  Deaths & 0.04 &   & 0.14 &   & 0.1 &   & 0.14 &   \\ 
  Asia &  &  & 0.47 &   &  &  & 2.42 &   \\ 
  Australia &  &  & 0.07 &   &  &  & 2.95 &   \\ 
  Europe &  &  & 0.99 & *  &  &  & 2.1 &   \\ 
  Latin America &  &  & 0.46 &   &  &  & 3.18 & *  \\ 
  North America &  &  & 1.39 & **  &  &  & 2.3 &   \\ 
  Adj. $R^2$ & -0.01 &   & 0.04 &   & -0.01 &   & 0.02 &   \\ 
  \hline
\multicolumn{1}{c}{}&\multicolumn{8}{c}{ \textbf{ High-impact}}\\\cline{2-9}
\multicolumn{1}{c}{}&\multicolumn{4}{c}{ \textbf{ English (n = 38)}}&\multicolumn{4}{c}{ \textbf{Spanish (n = 20)}}   \\
&\multicolumn{1}{c}{$\beta_{1}$} && \multicolumn{1}{c}{$\beta_{2}$}& &\multicolumn{1}{c}{$\beta_{1}$} && \multicolumn{1}{c}{$\beta_{2}$}& \\\hline
 Intercept & -12.61 & *** & -12.95 & *** & -18.03 & *** & -18.73 & *** \\ 
  Deaths & 0.97 & *** & 0.92 & *** & 1.33 & **  & 1.45 & **  \\ 
  Asia &  &  & 0.49 &   &  &  & -0.22 &   \\ 
  Australia &  &  &  &  &  &  &  &  \\ 
  Europe &  &  & 1.01 & *  &  &  & 0.88 &   \\ 
  Latin America &  &  & -0.21 &   &  &  &  &  \\ 
  North America &  &  & 1.72 & *  &  &  &  &  \\ 
  Adj. $R^2$ & 0.38 & *** & 0.48 & *** & 0.28 & **  & 0.50 & **  \\ 
   \hline
\end{tabular}
\caption{Results from regression analyses with logarithm of the maximum number of page views as dependent variable. The column for $\beta{1}$ corresponds to a model that only considers the number of deaths (log-transformed) as the independent variable, whereas $\beta{2}$ reports a model which considers log(deaths) and the airline region as independent variables. Significance codes: *** $<0.001$, ** $<0.01$, * $<0.05$.} 

 \label{table_regression}
\end{table*}

 If we split the data points into high- and low-impact events and recalculate the linear model separately for each regime, we see that the addition of the airline region in cases with high number of deaths increases the explanatory power of the regression. In both language editions, the proportion variance explained increases considerably. The explanatory power we obtain for the low-impact events, however, is negligibly small.

Based on the results of the categorical regression analysis including the location of the operating companies, one can estimate the relative level of attention paid to pairs of events from different regions on average. These ratios are reported in Table~\ref{table-equivalence}. For instance, controlling for the number of deaths, a North American event triggers about 50 times more attention among English Wikipedia readers compared to an African event. This ratio for North American versus European is about two. In Spanish Wikipedia however, a Latin American event triggers about 50 times more attention than an African and 5 times more than a North American event. 

\begin{table}[h]
\centering
\begin{tabular}{lcccccc}
  \multicolumn{1}{c}{} & \multicolumn{6}{c}{\textbf{ English Wikipedia}} \\ \cline{2-7}
 &Africa&Australia&Latin America&Asia&Europe&North America\\
Africa&1&2&2&6&26&47\\
Australia&&1&1&4&16&28\\
Latin America&&&1&4&16&28\\
Asia&&&&1&4&8\\
Europe&&&&&1&2\\
North America&&&&&&1\\
\vspace{.5cm}\\
  \multicolumn{1}{c}{} & \multicolumn{6}{c}{\textbf{ Spanish Wikipedia}} \\ \cline{2-7}
&Africa&Asia&North America&Australia&Europe&Latin America\\
Africa&1&5&10&10&16&48\\
Asia&&1&2&2&3&10\\
North America&&&1&1&2&5\\  
Australia&&&&1&2&5\\
Europe&&&&&1&3\\
Latin America&&&&&&1\\

\end{tabular}
\caption{ Death equivalence ratios based on the viewership data from English and Spanish Wikipedias. The matrix is calculated according to the coefficients reported on the upper part of Table \ref{table_regression}. For 6 different airline continents, the matrix shows the ratio of triggered attention, controlling for the number of deaths. For example, the attention given to events caused by a North American Airline in English Wikipedia is on average 2 and 47 times larger than to the events caused by  European and African companies respectively.  In Spanish Wikipedia, the level of attention given to events related to Latin America is 3 times larger than the European events, 5 times larger than North American, and 10 times larger than Asian events.}\label{table-equivalence}
\end{table}

\subsection{ Modeling attention decay}  

\label{sec:attdyn}

\begin{figure}[ht]
                \centering
                       \includegraphics[width=0.5\textwidth]{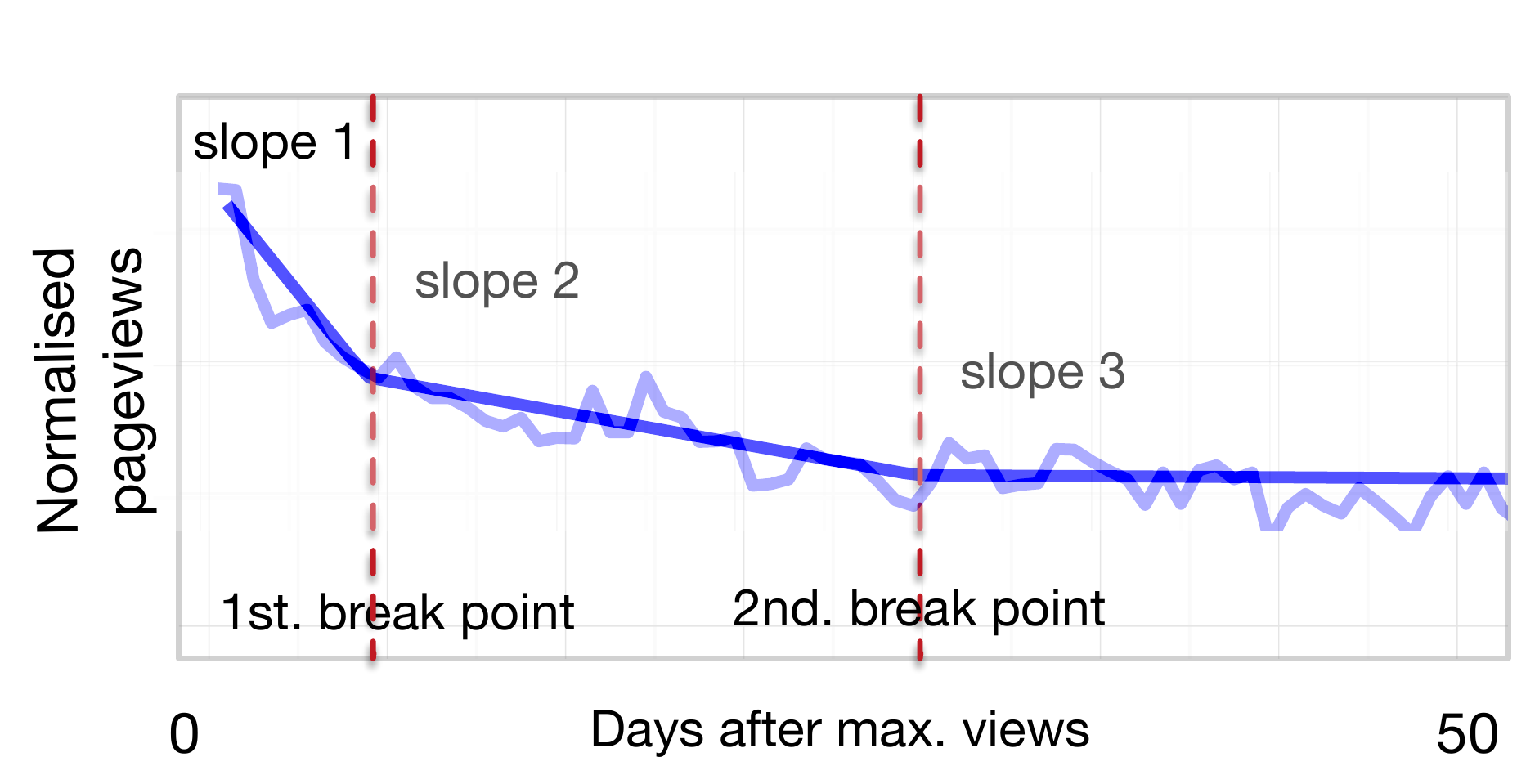}

  \caption{Typical example of the viewership time-series of a Wikipedia article related to an airplane crash fitted with segmented regression with two break points. The y-axis is in logarithmic scale.} 
  \label{fig_example_fitting}
\end{figure}

Now we focus on attention decay by analysing the viewership time-series after the event. After the initial boost in viewership, which in 73\% of the cases happens in less than 5 days after the date of the page creation, 
an exponential decay follows (see Figure~\ref{fig_example_fitting} for an example). This phenomenon occurs both due to the decay of novelty \cite{wu2007} as well as limitations in  human capacity to pay attention to older items in competition with newer ones~\cite{Parolo2015}. 

To model the attention decay, we use a segmented regression model with two break points to fit the normalized daily page-view counts in logarithmic scale (see Section \ref{sec:matmeth} for details). 
Figure~\ref{fig_example_fitting} shows a typical example of
the time series of the viewership of an article and the fit of the segmented regression model. 

The distributions of fit parameters are reported in Table~\ref{table_distributions}. These distributions confirm the assumptions that we make in developing our segmented regression model with two break points as well as similarities between the two language editions that we study. For instance, in both cases the half-life of the attention in the first phase and the detected position of the first break point show similar patterns.

\begin{figure}[ht]
\centering
 \includegraphics[width=0.8\textwidth]{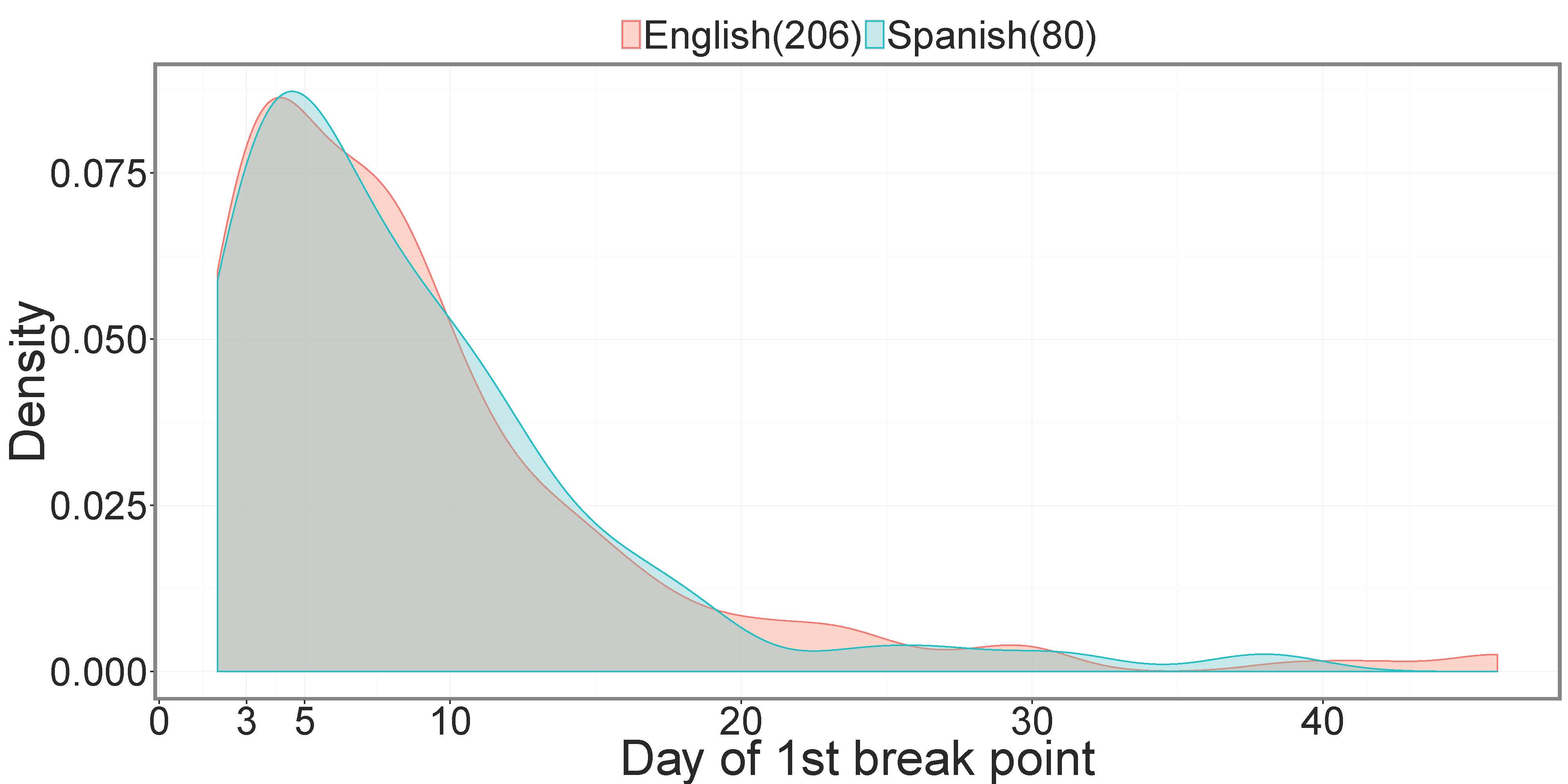}

 \caption{Distribution of the position of the first break point in number of days for a set of articles in English and Spanish Wikipedia. 
 }\label{fig_firstbp}

\end{figure}

 In Figure~\ref{fig_firstbp}, we show the distribution of the location of the first break point in larger scale. This parameter indicates the time span of the initial attention paid to the event. The first break point is localized around 3-10 days for both English and Spanish Wikipedia. 
 
\begin{figure}[ht]
\centering
 
\includegraphics[width=1\textwidth]{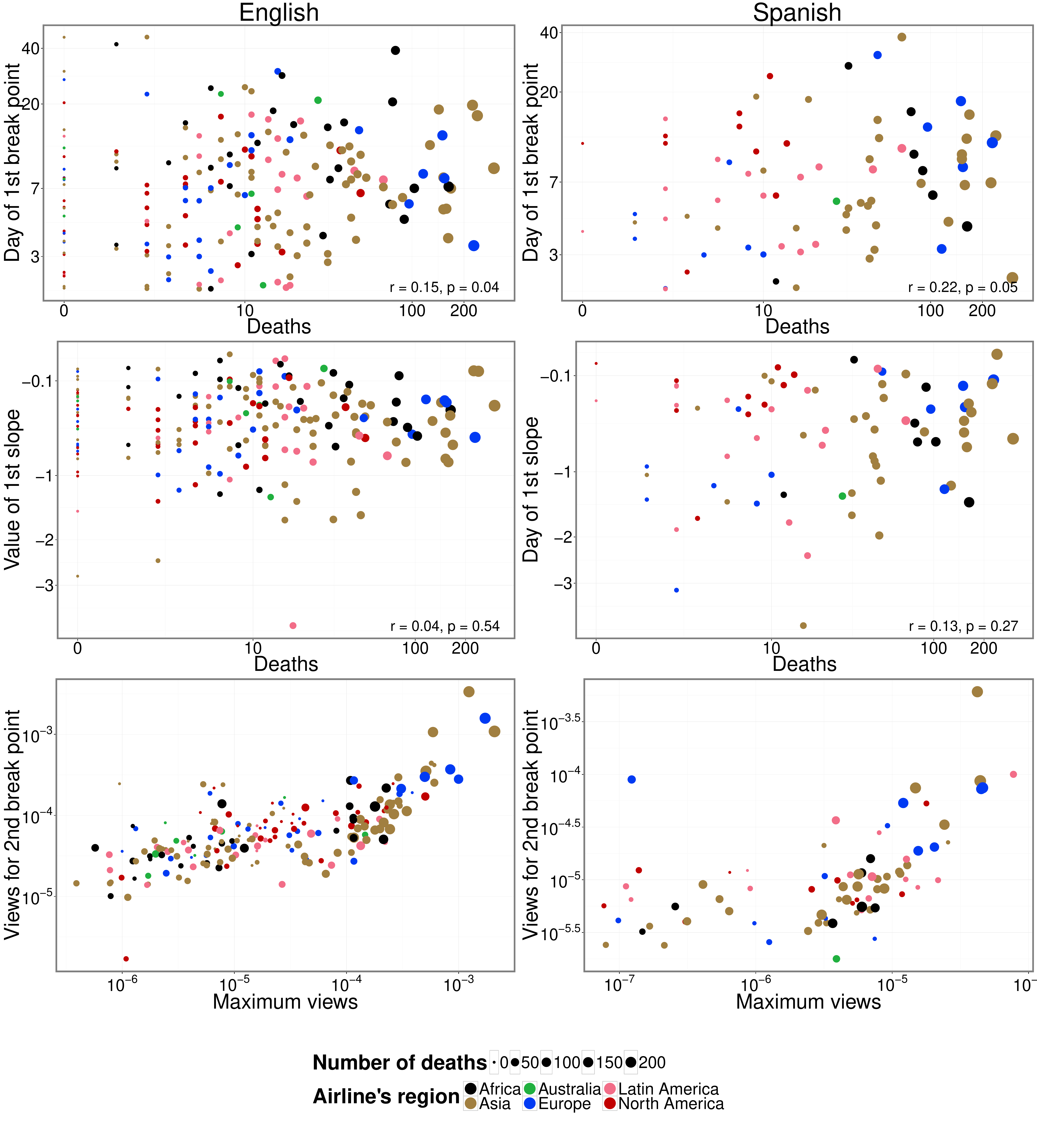} 
 \caption{Log-log scatter plots of model parameters against the number of deaths of each event: the first row shows the location of the first break point (days) versus the number of deaths, the second row shows the slope of the first segment versus the number of deaths, and the third row shows the intercept of the last segment versus the maximum daily page views. The four first plots report the Spearman's rank correlation coefficient and the corresponding p-value between the x and y axes.}\label{fig_first_death}
\end{figure}

 In Figure~\ref{fig_first_death} we consider other parameters that the best fit of the model assigns to each event. We observe that there is no significant correlation between the position and the value of attention at the first break point and the number of deaths, meaning that the rate of decay in attention and the first attention phase time span are independent of the impact of the event (upper and middle rows). However, in the lower row of the same figure we show that the relation between the level of attention at the second break point, which can be interpreted as the level of the long-lasting attention, and the immediate attention in the initial phase, is similar to what is observed in Figure~\ref{fig_regressions}, i.e., for low impact events, the long lasting attention is independent of the initial attention, whereas for high impact events, the initial attention is a good predictor of the long term attention to the event.

\section{Discussion and Conclusion}

We studied online attention to aircraft incidents and accidents using editorial and viewership data for the English and Spanish editions of Wikipedia. 
Overall, we found certain universal patterns. 

We found some differences in event coverage between the two languages but often, they can be attributed to the same underlying biases. For example, attention on English Wikipedia is more focused on events concerning North American and European airlines while attention on Spanish Wikipedia gives priority to Latin American airlines. English Wikipedia tends to cover more events in North America, while Spanish Wikipedia tends to cover more events in Latin America.

Our findings suggest that crashes of flights operated by North American companies, which mostly happened also in North America, receive higher publishing priority in English Wikipedia regardless
of the impact, while accidents from other locales, especially older accidents, are published later
and have to be more impactful to receive the same level of editorial attention.
Similar editorial biases in different contexts have been studied and reported before \cite{graham2014uneven,samoilenko2014distorted}. 
Although one can argue that English Wikipedia is mostly edited and used by North American users, previous research has shown that only about half of the editorial activity on English Wikipedia originates from North America \cite{yasseri2012b} and English should be considered as the {\it lingua franca} of Wikipedia \cite{kim2015}. Also note that the difference that we see within each Wikipedia language edition is consistent regardless of the language of the study and hence the origin of viewers. 
 
These biases in Wikipedia can be driven by the biases in mainstream media \cite{Adams1981}. Previous research has shown that a considerable dominance of references to Western media exists in Wikipedia \cite{fordget2013} and therefore, events of less importance for the Western media are more sparsely covered in Wikipedia. 
In the case of aircraft crashes, for example, in 1981, 10 people died in
the controversial flight {\it FAB 001} belonging to the Ecuadorian Air Force. It is a controversial flight because the former president of Ecuador
Jaime Rold\'{o}s was among the victims and the cause of the crash is still a mystery. Although there are articles in several languages in Wikipedia
covering the biography of Jaime Rold\'{o}s and the type of airplane used in the crash, there is no article equivalent to the specific flight that caused his death
and thus this case is missing in our dataset. The same happens for the flight that killed the former president of the Philipines Ram\'{o}n
Magsaysay or the Iraqi former president Abdul Salam Arif, among others.

In both languages, we observed two attention regimes for events -- low-impact regime, where the level of maximum attention is independent of the number of deaths and high-impact regime, where the airline region and the impact of the event significantly influence attention. In addition, focusing on the immediate attention to the event, we found that the time span and rate of the exponential decay (the slope of the fit to the first segment exemplified in the semi-log diagram of Figure~\ref{fig_example_fitting}) is independent of the impact of the event and the language of the article. The short span of attention that we observed (on the order of a few days) is in accordance with previous findings by other researchers \cite{wu2007,gleeson2014,Ciampaglia2015}.

Our study needs further generalization to include other type of events, such as natural disasters, political, and cultural events. Moreover, our analysis
has been limited to the English and Spanish editions of Wikipedia. Although these two are among the largest Wikipedia language editions, we might see variations in results studying attention patterns in different language editions.

\section{Materials and Methods}
\label{sec:matmeth}

\subsection{Data collection}

We collect data from Wikipedia using two main sources: the MediaWiki API and Wikidata. 
Wikidata\footnote{Using \url{https://cran.r-project.org/web/packages/WikidataR/index.html}.} is a Wikipedia partner project that aims to extract facts included in Wikipedia articles and fix inconsistencies across different editions~\cite{Lehmann2015}. 
Although content in Wikidata is still somewhat limited, the availability of such  structured information makes it easier for researchers to obtain data from a set of Wikipedia articles in a systematic way.

To complete the data missing from Wikidata, we automatically crawl Wikipedia infoboxes\footnote{Using \url{https://cran.r-project.org/web/packages/WikipediR/index.html}.} and collect features of events (see below).

We first focus on a set of articles classified as aircraft accidents or incidents in English Wikipedia, belonging to the categories \emph{Aviation accidents and incidents by country}  and \emph{Aviation accidents and incidents by year}, and their subcategories, which 
cover all airline accidents and incidents in different countries and throughout history available in Wikipedia. In total we obtain 1606 articles from which 1496 are specifically  about aircraft crashes or incidents
 (we discard articles of biographies, airport attacks, etc). From the 1496 articles, we obtain the following: date of the event, number of deaths, coordinates of the event, and airline region.

We extract all editorial information for the articles in the sample using the MediaWiki API. We extract the date when the article was created and alternative names for the article. We use the latter to merge all traffic 
statistics to the main title. Next, we extract all available articles in the same categories considered in English Wikipedia from Spanish and  follow the same procedure 
to extract the features of the articles in the Spanish edition. In total, we obtain 525 articles in Spanish Wikipedia from which 488 are about aircraft incidents or accidents.

Finally, we extract the daily traffic to the articles in English and Spanish from the Wikipedia pageview dumps\footnote{\url{https://dumps.wikimedia.org/other/pagecounts-raw/}} through a third party interface.\footnote{\url{http://stats.grok.se}}

\subsection{Data analysis}

To control for the changes in the overall popularity of Wikipedia, we normalize the viewership counts by the overall
 monthly traffic to Wikipedia.\footnote{The data are obtained from \url{https://stats.wikimedia.org/EN/Tablespage-viewsMonthlyCombined.htm}}
To numerically model attention dynamics, we apply segmented regression analysis to viewership data during 50 days after the first pick due to the occurrence of the event. We use segmented regression as implemented in the R package ``segmented''. \footnote{We use the R package \emph{segmented}: \url{https://cran.r-project.org/web/packages/segmented/}} Segmented regression models are models where the relationship between the response and one or more explanatory variables are piecewise linear, represented by two or more straight lines connected at values called breakpoints~\cite{Muggeo2008}. To find those breakpoints, the algorithm first fits a generic linear model then fits the piecewise regression through an iterative procedure that uses starting break point values given by us at the beginning. In our specific case, three piecewise regressions are fit in each iteration and the two break point values are updated accordingly as to minimize the gap~$\gamma$~between the segments. The model converges when the gap between the segments is minimized. We refer the reader to the paper by Muggeo~\cite{Muggeo2008} for a detailed explanation. Additionally, the package description explains that bootstrap restarting is used to make the algorithm less sensitive to starting values.

Although alternative approaches could be undertaken to model nonlinear relationships,
for instance via splines, the main appeal of the segmented model lies in its simplicity and the interpretability of the parameters.

We have chosen  two break points (three segments) for the analysis but our main results are robust against changing this number (see Figure \ref{fig_dist_1stbp} in the Appendix). This choice is informed by previous research that identifies three phases in the evolution of collective reactions to events: communicative interaction, floating gap, and cultural memory (stabilization phase)~\cite{Pentzold2009}.

We find that most of the events are fitted well, with high adjusted $R^2$ (average 0.84 for English and 0.80 for Spanish). However, in some cases, this model is not able to capture the overall dynamics, mostly due to
 secondary shocks driven by new triggering factors that are too close to the event, e.g., the discovery of the corresponding airplane black box or other related newsworthy events.

\begin{table}[t]
\small
\begin{center}
{

  \includegraphics[width=1\textwidth]{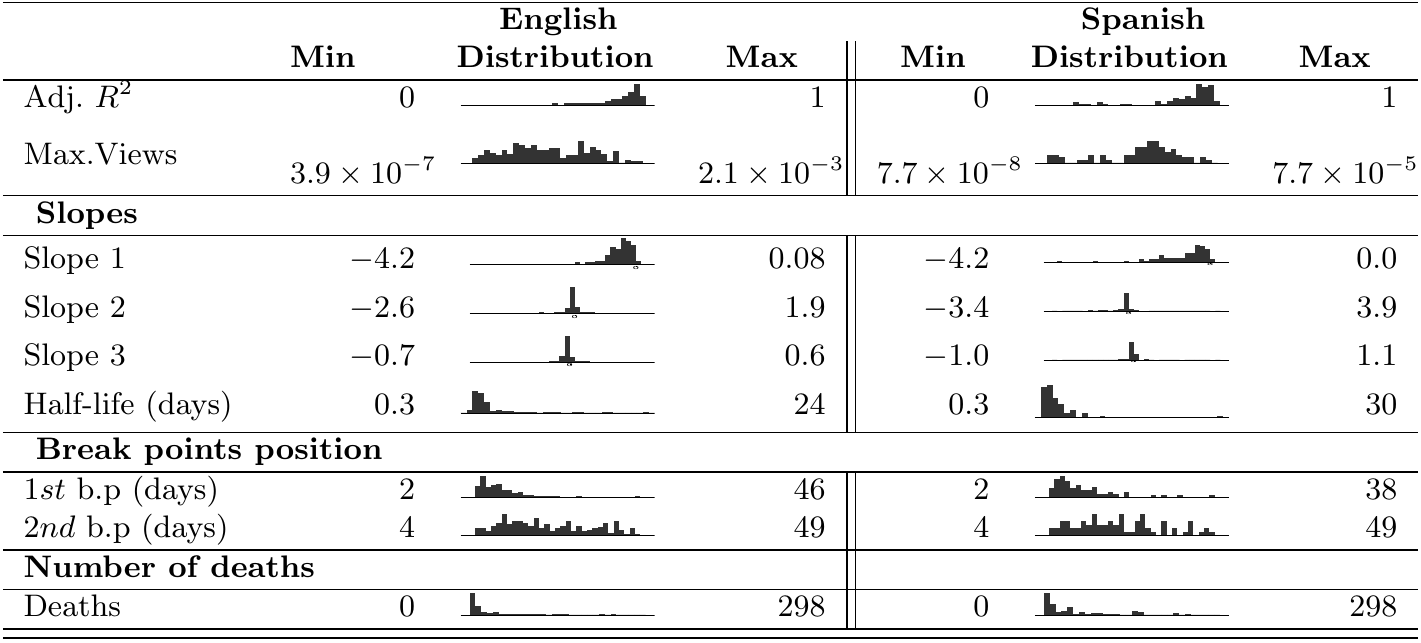}  
}
\end{center}
\caption{The distribution of normalized maximum daily views of each article and Adj.~$R^2$ of the segmented regressions as well as the distribution of the model parameters, calculated half-life (reverse of the absolout value of the slope), and the number of deaths for each event. All distributions are based on 206  and 80 observations for English (En) and Spanish (Sp) Wikipedias.}
\label{table_distributions}
\end{table}

\section*{Data Accessibility}
The datasets supporting this article have been uploaded to DRYAD system and is available via https://www.doi.org/10.5061/dryad.34mn3.

\section*{Competing interests}
The authors declare no competing interests.

\section*{Authors' contributions}
RG-G collected and analysed the data, participated in the design of the study, and drafted the manuscript; MT participated in the design of the study and helped draft the manuscript; TY conceived, designed, and coordinated the study, and helped draft the manuscript. All authors gave final approval for publication.



\section*{Funding}
This research is part of the project \emph{Collective Memory in the Digital Age: Understanding Forgetting on the Internet} funded by Google. 

\sloppy
\emergencystretch 1.5em
\bibliographystyle{vancouver}
\bibliography{main} 
%
%
%
%
%
%
\appendix
\section{Appendix}

\begin{figure}[ht]
\centering
 \subfloat[]{
\includegraphics[width=1\textwidth]{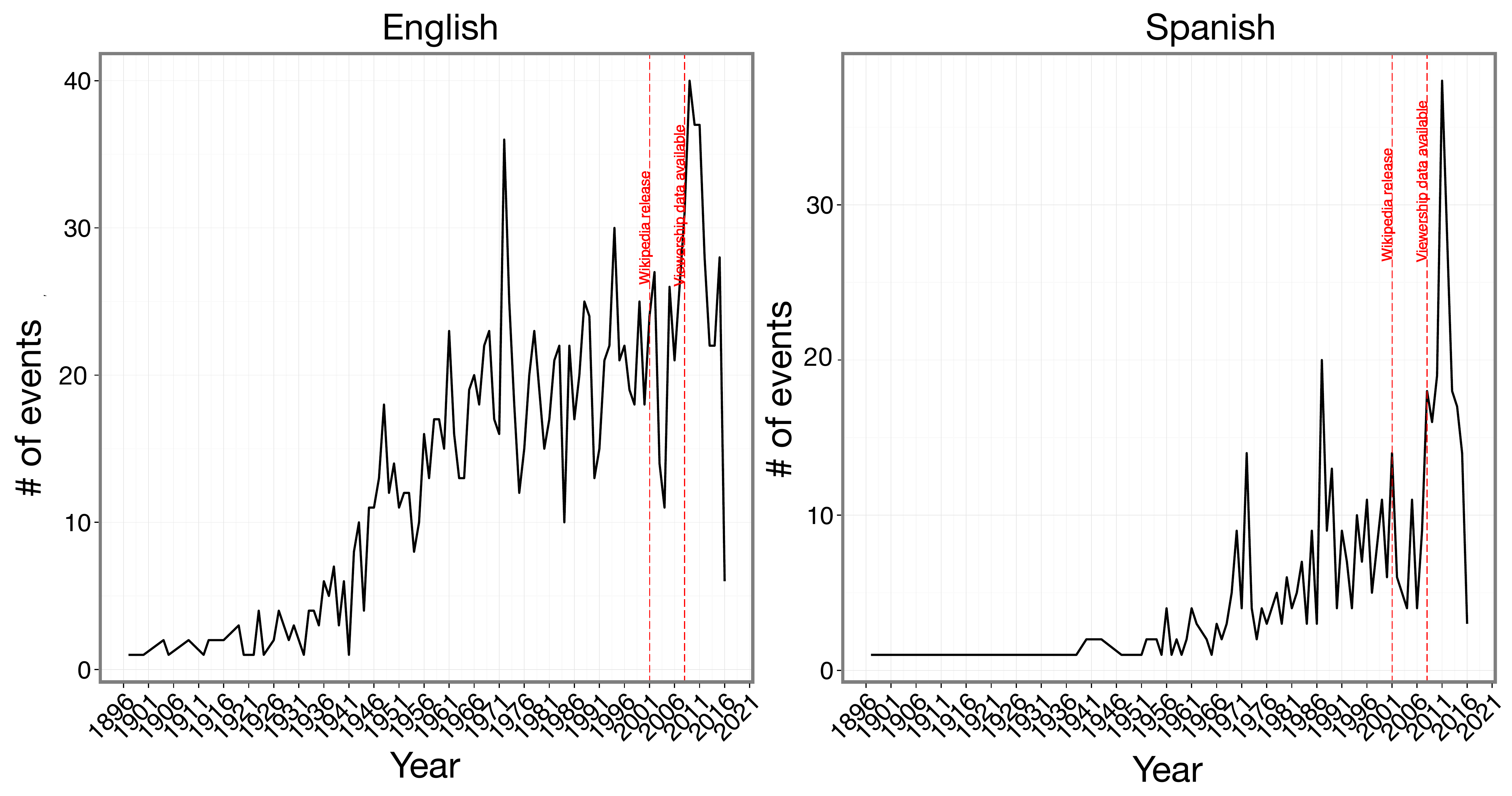} }
 \caption{The number of aircraft incidents and accidents per year reported in English and Spanish Wikipedia.}\label{fig_events_per_year}
\end{figure}

\begin{figure}[ht]
\centering
 \subfloat[]{
\includegraphics[width=1\textwidth]{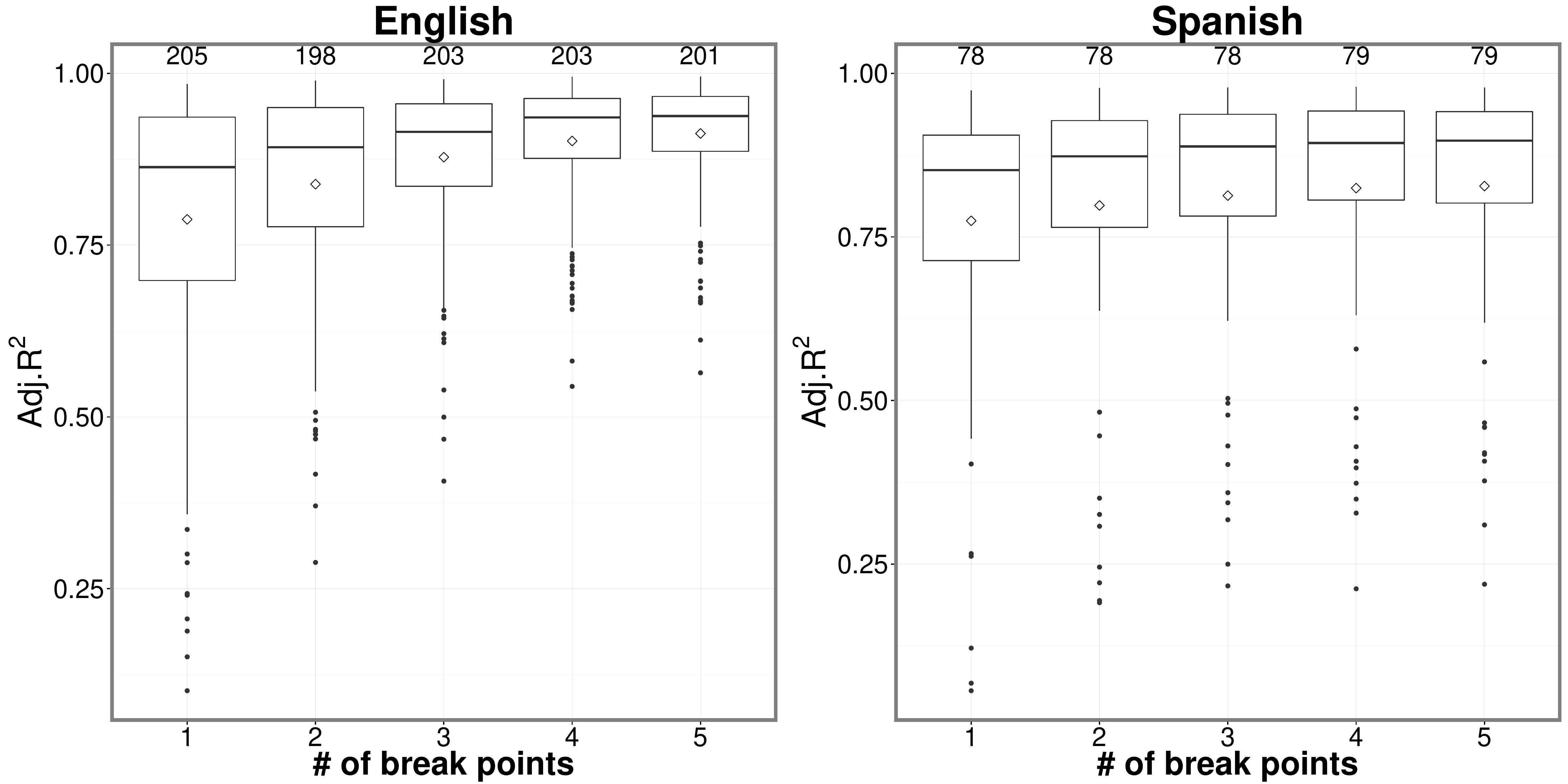} }
 \caption{Boxplot of the variance explained ($Adj.R^2$) of the viewership time series (up to 50 days after the event) of Wikipedia articles for different number of break points. The numbers at the top represent the total count of data points for each model. }\label{fig_boxplot}
\end{figure}

\begin{figure}[ht]
\centering
 \subfloat[]{
\includegraphics[width=1\textwidth]{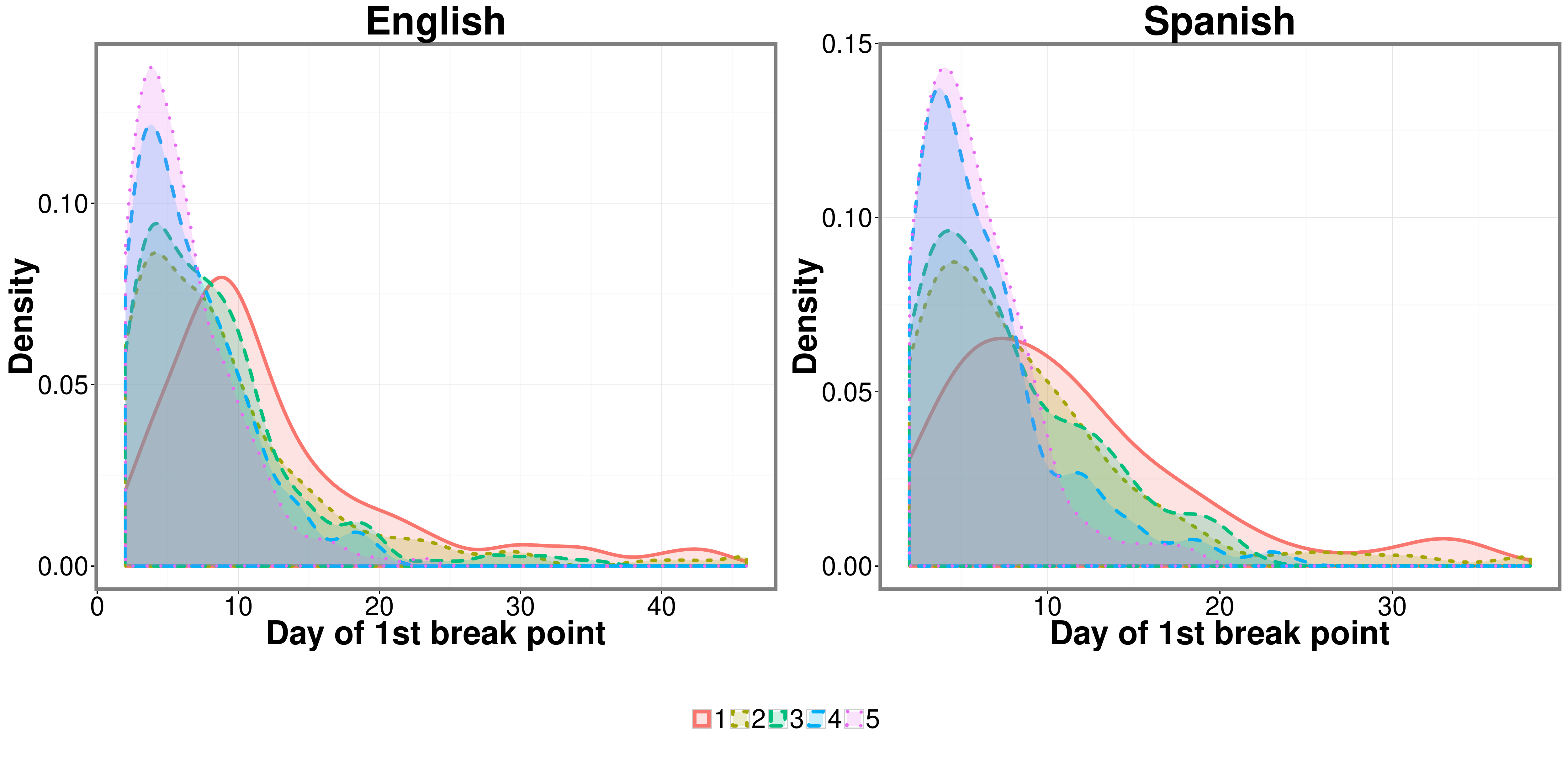} }
 \caption{Distribution of the location of the first break point (days) for segmented regressions with different number of break points.}\label{fig_dist_1stbp}
\end{figure}

\end{document}